\documentclass{aa}
\usepackage[sort]{natbib}
\bibpunct{(}{)}{;}{a}{}{,}  
\usepackage{graphicx}
\usepackage{txfonts}

\def\MBot{{M$_\mathrm{B_T^0}$}}
\def\N13{{N$_\mathrm{13}$}}
\def\MK5{{M$_\mathrm{K_{5}}$}}
\def\mqas{{mag arcsec$^\mathrm{-2}$}}

\begin{document}
\title{Statistics of young starforming complexes in spiral galaxies using NIR
  photometry \thanks{Based on observations collected at the European Southern
    Observatory, Chile; program: ESO 68.A-0085 and 69.A-0021}} 
\subtitle{}
\titlerunning{Young starforming complexes in spiral galaxies}
\author{P.~Grosb{\o}l\inst{1} \and H.~Dottori\inst{2}}
\offprints{P.~Grosb{\o}l}
\institute{European Southern Observatory,
  Karl-Schwarzschild-Str.~2, 85748 Garching, Germany \\
  \email{pgrosbol@eso.org}
\and
  Instituto de F\'{i}sica, Univ. Federal do Rio Grande do Sul,
  Av. Bento Gon\c{c}alves~9500, 91501-970 Porto Alegre, RS, Brazil \\
  \email{dottori@ufrgs.br}
}

\date{Received ??? / Accepted ???}

\abstract{} {Very young stellar clusters and cluster complexes may
  be embedded in dust lanes along spiral arms in disk galaxies and escape
  detection in visual bands.  Observations in the near-infrared K-band offer
  an almost unbiased view of such clusters or complexes due to the
  small attenuation by dust at this wavelength.  The objective is to determine
  their population size, absolute K-band magnitude distribution above the
  limiting magnitude imposed by the data, and location relative to the spiral
  pattern in disk galaxies.}
{All slightly extended sources were identified on deep K-band maps of 46
  spiral galaxies reaching at least K=20.3 \mqas\ at a signal-to-noise level
  of 3.  The galaxies had inclination angles $<65\degr$ and linear
  resolutions $<100$~pc with seeing better than $1\arcsec$. The sample
  includes both barred and normal spirals with a wide spread in types. We also
  analyzed J- and  H-band colors for 4 galaxies for which such images were
  available.  An apparent magnitude limit of K = 19~mag was used for the
  sources analyzed in order to avoid marginal detections.  Furthermore, we
  derived the source distributions of magnitudes and relative locations with
  respect to the spiral patterns. }
{Almost 70\% (15/22) of the grand-design spiral galaxies show significant
  concentration of bright K-band knots in their arm regions corresponding to
  30\% (15/46) of the full sample.  Color-color diagrams for the 4 spirals
  with JHK photometry suggest that a significant fraction of the
  diffuse sources found in the arms are complexes of young stellar
  clusters with ages $<$10~Myr and reddened with several magnitudes of visual
  extinction. The brightest knots reach an absolute K-band magnitude
  M$_\mathrm{K}$ of -15.5~mag corresponding to stellar clusters or
  complexes with total masses up to at least 10$^5$~M$_{\odot}$.  Brightest
  magnitude and number of knots correlate with the total absolute magnitude of
  the host galaxy.  More knots are seen in galaxies with high far-infrared
  flux and strong two-armed spiral perturbations.  The bright knots constitute
  up to a few percent of the total K-band flux from their parent galaxy and
  account for a star formation rate of $\sim$1~M$_{\odot}$ yr$^{-1}$ for the
  brightest grand-design spiral galaxies.  } {}
\keywords{galaxies:~spiral -- galaxies:~structure -- galaxies:~star~clusters --
  infrared:~galaxies -- techniques:~photometric}
\maketitle
\section{Introduction}
The present star formation rate (SFR) in the local Universe is important as a
part of the history of galaxy build-up and as a means of comparison for the
study of star formation at higher redshifts.  For spiral galaxies, significant
variation is observed as a function of Hubble type and luminosity class.
Extinction by dust in spiral arms makes it difficult to get a full census from
observations in visual bands, whereas near-infrared (NIR) colors offer a much
higher transparency. 

In a study of the NIR K-band images of spiral galaxies, \citet{gp98} noticed
that several grand-design spirals have bright knots along their arms,
suggesting that such knots are related to the spiral structure. Young stellar
clusters may expel gas left over after their main star formation phase and may
then experience violent relaxation \citep{bastian06}.  This could lead to
their destruction on time scales of a few tens of Myr \citep{goodwin06}.  The
knots are marginally resolved, suggesting sizes of less than 100 pc.  Unbound
stellar clusters will exceed this size due to diffusion of stellar orbits
within 50~Myr \citep{wielen77}. These points indicate that the knots are
dynamically young.

\citet{patsis01} studied such knots in two galaxies using narrow-band filters
in the K-band and concluded that the major contribution to their K-band fluxes
comes from continuum radiation.  In many cases these knots are embedded in
dust lanes and invisible in visual bands. Taking advantage of the mutual
alignment of eight of these knots along the southern arm of NGC 2997,
\citet{grosbol06} obtained spectroscopy in the J- and K-bands detecting
hydrogen Br$_\gamma$ in emission, undoubtedly demonstrating that the knots are
HII regions deeply enshrouded in dust, confirming the results previously
suggested by \citet{patsis01}. They also explored the relative position of the
knots with respect to the ridge of the K-band arm to derive kinematic
parameters of the density wave to constrain the star formation regime and ages
of the knots.

Their absolute luminosity and Br$_\gamma$ emission suggest they are very
young, massive stellar complexes that would host a significant fraction of the
heaviest stars formed in a galaxy.  Thus, the statistics of them in spiral
galaxies may imply SFR of stars on the upper part of the initial mass function
(IMF) and rate of supernovae type II in such systems.  The low attenuation by
dust in the K-band makes it possible to obtain almost complete samples while
many clusters and complexes may be missed on visual images due to heavy
obscuration in dust lanes.  To estimate internal mean physical properties of
these NIR starforming knots, we used the theoretical models starburst99 by
\citet[ hereafter SB99]{leitherer99}. The rate of star formation, derived from
NIR data, may complement or correct the one derived from the observations of
optical HII regions, as done by SINGG \citep{meurer06, hanish06}. Indeed, due
to the dust absorption, few of the NIR knots would contribute significantly to
the H$_{\alpha}$ emission line, used by SINGG to estimate the star formation
rate in their sample of galaxies.

Thus, the knots we are studying in this paper may constitute the first phases
of the star formation that escape detection in optical bands, or might well be
the dusty, densest part of large HII region complexes. We present statistics
of such knots detected on K-band images of 46 spiral galaxies.  For the four
galaxies observed in the J-, \mbox{H-,} and K-bands, we also compared their
colors with simple SB99 models using different star formation scenarios.

\section{Data and reductions}
The current sample of galaxies was selected from two data sets containing
deep, high-resolution K-band images of nearby spiral galaxies.  The first set,
designated A, was observed to study the morphology of bar and spiral
perturbations in disk galaxies through K-band imaging \citep{grosbol04}, while
the second set (B) with both H- and K-band maps included galaxies in which
supernovae Ia had recently been observed.  Some galaxies in the second set
also had J-band images and were designated with (C).  All observations were
made with SOFI at the NTT telescope at ESO, La Silla, using a Rockwell
Hg:Cd:Te 1024$\times$1024 detector with a pixel size of 0.29\arcsec.  The
preliminary sample was 95 galaxies, selecting only targets with inclination
angles $i < 65$\degr\ and seeing $< 1$\arcsec.

General data for the galaxies were taken from the Third Reference Catalog
\citep[ hereafter RC3]{rc3} such as types, systemic velocities, total
magnitudes, and radio fluxes.  Distances were computed from the velocity
V$_\mathrm{GSR}$, corrected to the Galactic Standard of Rest, using a Hubble
constant of 75 km sec$^{-1}$ Mpc$^{-1}$.

The final sample was restricted to galaxies with distances D $> 5$~Mpc, to
avoid galaxies in which individual stars could be resolved, absolute blue
magnitudes M$_\mathrm{B} < -19.0$~mag, to exclude dwarf galaxies, surface
brightness at a signal-to-noise (S/N) of 3 for the K-band maps $\mu_\mathrm{3}
> 20.0$~\mqas~, and a linear resolution $\Delta$s $<100$~pc.  The 46 spiral
galaxies in the sample are listed in Table~\ref{tab1}, which also gives their
Hubble type and luminosity class. The Hubble types listed are taken from RC3
where the two first letters give the main morphology with SA for ordinary
spirals, SB for barred systems, and SX for intermediate or weakly barred
galaxies.  The third character indicate the bar/spiral interface, whereas the
final number gives the numeric type T going from (1) for early-type spirals to
(9) for late-type systems.  The sample is slightly biased against barred
spirals with 18~SA, 21~SX, and only 7~SB galaxies due to the definition of the
two original data sets.  Eight galaxies (NGC~1068, NGC~3351, NGC~3368,
NGC~3627, NGC~4321, NGC~4535, NGC~4579, and NGC~5247) are in common with the
BIMA SONG survey \citep{helfer03}, while 6 galaxies (NGC~1566, NGC~3351,
NGC~3621, NGC~3627, NGC~4321, and NGC~4579) are also in the Spitzer SINGS
sample \citep{kennicutt03}.

\begin{table}
  \caption[]{List of galaxies.} 
 \label{tab1}
 \begin{tabular}{llcccrrr}
\hline\hline
  Galaxy & Type &  \multicolumn{1}{c}{L} & 
  Set & FWHM & \multicolumn{1}{c}{$\mu_\mathrm{3}$} &
  \multicolumn{1}{c}{D} & \multicolumn{1}{c}{$\Delta$s} \\ \hline
   {NGC 157}  & SXT4 &  1.8 &    A &  0.7 & 21.1 & 23.1 &   80 \\
   {NGC 210}  & SXS3 &  1.1 &    A &  0.9 & 20.7 & 22.4 &   96 \\
   {NGC 289}  & SBT4 &  2.2 &    B &  1.0 & 20.8 & 21.5 &   99 \\
   {NGC 578}  & SXT5 &  2.4 &    A &  0.6 & 20.7 & 21.5 &   61 \\
   {NGC 685}  & SXR5 &  4.0 &    A &  0.8 & 20.7 & 16.6 &   65 \\
   {NGC 895}  & SAS6 &  1.9 &    A &  0.6 & 20.8 & 30.6 &   95 \\
   {NGC 1068} & SAT3 &  2.3 &    A &  0.9 & 20.7 &  8.2 &   34 \\
   {NGC 1084} & SAS5 &  3.1 &    B &  0.8 & 20.9 & 18.5 &   76 \\
   {NGC 1087} & SXT5 &  5.5 &    A &  0.8 & 20.7 & 20.3 &   74 \\
   {NGC 1187} & SBR5 &  2.1 &    B &  0.9 & 20.8 & 17.7 &   73 \\
   {NGC 1255} & SXT4 &  3.3 &    A &  0.8 & 21.0 & 21.5 &   85 \\
   {NGC 1300} & SBT4 &  1.1 &    B &  0.8 & 20.7 & 19.9 &   80 \\
   {NGC 1350} & SBR2 &  3.0 &    B &  0.7 & 20.7 & 23.6 &   82 \\
   {NGC 1357} & SAS2 &    - &    A &  0.5 & 20.7 & 25.8 &   66 \\
   {NGC 1365} & SBS3 &  1.3 &    B &  0.8 & 21.1 & 20.5 &   85 \\
   {NGC 1371} & SXT1 &    - &    A &  0.8 & 21.1 & 18.3 &   67 \\
   {NGC 1398} & SBR2 &  1.1 &    B &  0.8 & 20.5 & 17.4 &   64 \\
   {NGC 1425} & SAS3 &  3.2 &    B &  1.0 & 20.9 & 18.6 &   89 \\
   {NGC 1566} & SXS4 &  1.7 &    B &  0.9 & 20.9 & 17.6 &   76 \\
   {NGC 2775} & SAR2 &    - &    A &  0.6 & 20.8 & 16.3 &   50 \\
   {NGC 2855} & SAT0 &    - &    A &  0.7 & 20.9 & 23.0 &   81 \\
   {NGC 2935} & SXS3 &  2.1 &    B &  0.6 & 20.4 & 27.6 &   85 \\
   {NGC 2997} & SXT5 &  1.6 &    A &  0.8 & 20.9 & 11.6 &   43 \\
   {NGC 3351} & SBR3 &  3.3 &    B &  0.6 & 20.4 &  9.0 &   28 \\
   {NGC 3368} & SXT2 &  3.4 &    B &  0.6 & 20.4 & 10.6 &   33 \\
   {NGC 3370} & SAS5 &  3.4 &    B &  0.6 & 20.4 & 16.0 &   45 \\
   {NGC 3389} & SAS5 &  4.2 &    B &  0.7 & 20.4 & 16.0 &   51 \\
   {NGC 3621} & SAS7 &  5.8 &    B &  0.7 & 20.5 &  7.0 &   22 \\
   {NGC 3627} & SXS3 &  3.0 &    B &  0.7 & 20.5 &  8.6 &   30 \\
   {NGC 3810} & SAT5 &  2.4 &    B &  0.9 & 20.5 & 12.2 &   55 \\
   {NGC 4030} & SAS4 &  1.6 &    A &  0.9 & 20.7 & 18.0 &   75 \\
   {NGC 4321} & SXS4 &  1.1 &    B &  0.9 & 20.5 & 20.5 &   94 \\
   {NGC 4535} & SXS5 &  1.6 &    B &  0.6 & 20.5 & 25.2 &   71 \\
   {NGC 4579} & SXT3 &  3.1 &    B &  1.0 & 20.5 & 19.6 &   94 \\
   {NGC 4941} & SXR2 &  3.4 &    A &  1.0 & 20.8 & 13.6 &   65 \\
   {NGC 4965} & SXS7 &  4.7 &    A &  0.7 & 20.6 & 28.3 &   98 \\
   {NGC 4981} & SXR4 &  3.4 &    A &  0.7 & 20.3 & 21.3 &   75 \\
   {NGC 5085} & SAS5 &  2.7 &    C &  0.8 & 20.6 & 24.4 &   89 \\
   {NGC 5134} & SAS3 &  4.7 &    B &  0.8 & 20.6 & 21.8 &   87 \\
   {NGC 5247} & SAS4 &  1.8 &    C &  0.7 & 20.6 & 16.7 &   60 \\
   {NGC 5364} & SAT4 &  1.1 &    C &  0.7 & 20.6 & 16.2 &   57 \\
   {NGC 5668} & SAS7 &  4.3 &    B &  0.8 & 20.6 & 21.1 &   77 \\
   {NGC 5861} & SXT5 &  2.2 &    C &  0.7 & 20.6 & 24.8 &   88 \\
   {NGC 6384} & SXR4 &  1.1 &    B &  0.8 & 20.5 & 23.8 &   87 \\
   {NGC 7213} & SAS1 &    - &    A &  0.9 & 20.3 & 23.6 &   99 \\
   {NGC 7418} & SXT6 &  3.5 &    A &  0.8 & 20.7 & 19.3 &   78 \\
  \hline   
\end{tabular}
\end{table}

All observations were done with the same template, which included a jitter
pattern on the target with offsets of around 10\arcsec\ interleaved with sky
exposures around 10\arcmin\ from the galaxy. Sky fields were explicitly
selected not to contain bright stars.  Typical exposure times on target were
4$^\mathrm{m}$, 4$^\mathrm{m}$, and 10$^\mathrm{m}$ for the J-, \mbox{H-,} and
K-bands, respectively, whereas some galaxies in data set (A) were exposed
longer. For a few galaxies with major axes larger than 5\arcmin\ several SOFI
fields were mosaiced together. The reductions of the frames followed the
procedure described by \citet{grosbol04} including determination of sky
projection parameters. Depth, expressed as $\mu_\mathrm{3}$, and seeing of the
final maps are given in Table~\ref{tab1}.

Photometric calibration was done by observing standard stars in the list of
\citet{persson98} several times during each night.  The error of the K-band
zero-point for galaxies from data set (A) was around 0.07~mag, while the
remaining galaxies had errors of 0.03~mag.  Typical errors for the H- and
J-band were estimated to 0.03~mag. For the four galaxies with J-, \mbox{H-,}
and K-band observations, the location of foreground stars in the (J-H)--(H-K)
diagram can be used to check zero-point errors as they have a characteristic
$\Gamma$ shape with a relative narrow range (see Fig.~\ref{all-ccd}).
Comparing their location with that of the local main-sequence and giant stars,
color offsets were estimated to $\Delta$(J-H) $\approx$ -0.07~mag and
$\Delta$(H-K) $\approx$ 0.06~mag.  These corrections were applied to the
colors, while the K-band magnitudes were not modified as the origin of the
color offsets are unclear (possibly a zero-point error in the H-band
photometry).  The positional errors were on average 0.5\arcsec\ and the actual
detector pixel size estimated to be 0.288\arcsec.

\begin{figure}
  \resizebox{\hsize}{!}{\includegraphics{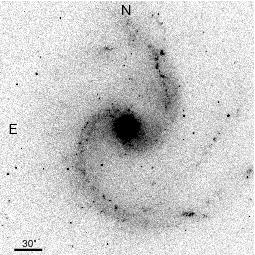}}
  \caption[]{Direct K-band image of NGC~5247 in intensity scale where the bar
  indicates scale.}
  \label{n5247k}
\end{figure}

\section{Detection of sources}
\label{sec-sources}
The maps of the galaxies have linear resolutions in the range of 20-100~pc,
which in most cases are not sufficient to resolve individual stellar clusters.
The majority of sources detected in the images are therefore more likely to be
associated to complexes of clusters rather than single discrete clusters.  To
check this issue, the Hubble Space Telescope archive was searched for ACS/WFC
frames of the galaxies taken with the F814W filter.  Reasonable deep exposures
in 7 galaxies (i.e. NGC~1068, NGC~1300, NGC~3368, NGC~3370, NGC~3621,
NGC~3810, and NGC 6384) were identified with linear scales in the range of
2-5~pc pixel$^{-1}$.  The comparison of our K-band images with the ACS/WFC,
with more than 10 times better resolution, suggests that the non-stellar
sources detected are a mixture of both individual stellar clusters and more
complex aggregates of starforming regions sometimes associated with ionized
gas.  Several objects seen on the K-band frames had no obvious counterpart on
the ACS/WFC frames possibly due to the significant extinction by dust in the
F814W filter.

\begin{figure}
  \resizebox{\hsize}{!}{\includegraphics{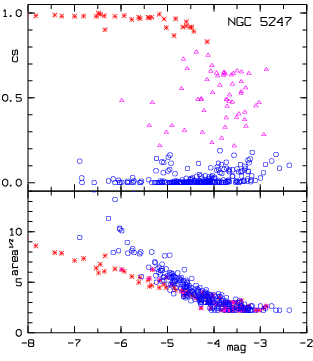}}
  \caption[]{Distribution of sources found by {\sl SExtractor} on the K-band
    image of NGC~5247. Upper panel shows the {\it class\_star} classifier, cs,
    as a function of internal aperture magnitude. Objects are indicated with
    ($\ast$) for 0.8$<$cs, ($\triangle$) for 0.2$<$cs$\leq$0.8, and ($\circ$)
    for cs$\leq$0.2. Lower panel displays the square root of the object area
    in pixels as a function of magnitude using the same symbols.}
  \label{sext}
\end{figure}

 The {\it SExtractor} program v.2.5 \citep{bertin96} was applied to the frames
 to identify all significant sources in the fields. A detection threshold of
 2.5 times local noise and a minimum area for objects of 5 pixels were used.
 No internal digital filtering was selected, and the background mesh was
 reduced to 16 so as to better follow variation in the arm regions of the
 galaxies.  Location and basic photometric parameters for the sources as
 detected on the K-band images were saved with the galaxy--star classifier
 {\sl class\_star}. Aperture photometry was then performed on all frames using
 the positions derived from {\it SExtractor} and a diameter of 3.0\arcsec.
 This procedure gave significant smaller scatter in the color indices compared
 to an independent search on frames in the other colors since {\it SExtractor}
 often yields slightly different source coordinates for each frame.  Aperture
 magnitudes were used since isophotal magnitudes depend too strongly on
 the exact inclusion of pixels within the isophote on different color maps and
 therefore display a wider spread in color indices.

NGC~5247 (see Fig.~\ref{n5247k} for a K-band image) was selected to illustrate
the results since all three NIR colors were available, and its morphology
provides a clean view of its spiral structure. {\sl SExtractor} found a total
of 337 sources, which fulfilled the criteria on the K-band image.  The
galaxy--star classifier {\sl class\_star} (denoted by cs), which ranges from
0.0 for diffuse, extended sources to 1.0 for stellar type objects, is shown as
a function of the internal aperture magnitude in the upper panel of
Fig.~\ref{sext}, while the lower one displays the square root of the number of
pixels included in the source.  A linear relation between magnitude and square
root of the area is expected for stellar type objects with Gaussian profiles
(as seen in Fig.~\ref{sext}). More diffuse sources have larger areas for the
same magnitude and scatter above the relation for stars.

\begin{figure}
  \resizebox{\hsize}{!}{\includegraphics{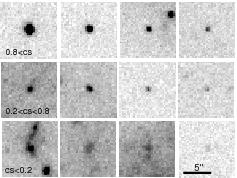}}
  \caption[]{ Cutouts of sources from the K-band image of NGC~5247 in
    Fig.~\ref{n5247k}.  The upper row displays sources classified as star-like
    objects with 0.8$<$cs. The middle row shows compact sources with
    0.2$<$cs$\leq$0.8 while the lower row illustrates diffuse sources with
    cs$\leq$0.2.  Targets in all rows are arranged with the fainter sources
    to the right. The bar in the lower right corner indicates the scale
    for all subframes.}
  \label{cutoufs}
\end{figure}

Depending on the galaxy--star classifier, the sources were divided into three
categories, namely: star-like objects with 0.8$<$cs, compact sources in the
range 0.2$<$cs$\leq$0.8, and diffuse targets with cs$\leq$0.2.  These are
illustrated on Fig~\ref{cutoufs} where cutouts around typical sources in the
NGC~5247 field (see Fig~\ref{n5247k}) are presented.  The subframes are
arranged in three rows, one for each category, and show brighter sources to
the left.  The linear scale of NGC~5247 is 81~pc arcsec$^{-1}$ while the
resolution corresponds to 60~pc.  This suggests that most of the extended
sources associated to the galaxy are complexes of clusters and/or HII regions
rather than individual clusters.

The sources detected in the fields can generally be placed in three groups,
namely objects associated to the galaxies, foreground stars, and background
galaxies.  To deal with foreground star contamination, we used the synthetic
model of structure and evolution of the Milky Way \citep{robin03, robin04},
which allows an estimate of the amount of stars up to a certain brightness
limit expected in a given sky direction.  We crosschecked this number for each
galaxy with the number of star-like objects (0.8$<$cs) in the same field
furnished by {\sl SExtractor}.  Since the galaxies in our sample are at
reasonably high galactic latitudes, both numbers agree closely when adopting a
mean Milky Way extinction model. Thus, we concluded that foreground stars
could safely be rejected by relying on the {\sl SExtractor} classification.

The number of background galaxies expected for a given field was estimated
with the program {\sl GalaxyCount} developed by \citet{ellis06}. Within a
galactic disk of 10~arcmin$^2$, 44$\pm$9 galaxies brighter than K~=~19 mag are
expected.  The brighter part of the stellar cluster is less contaminated
because only 6$\pm$2 galaxies with K $<$ 17~mag are estimated. These values
are average estimates and do not take the existence of clusters of background
galaxies into account in the field.  This was checked by visual inspection of
all K-band frames.  Background clusters were found in the fields of NGC~1255,
NGC~2775, NGC~3389, and NGC~5668 but in all cases well outside the disk of the
galaxies.  To reduce the possible contamination by background galaxies, only
sources within the disk of the galaxies were considered. The disk radius was
estimated on the K-band frames as the higher value of either the outer extent
of the spiral pattern or the radius within which 90\% of galaxy's K-band
luminosity was emitted.

The globular cluster systems of the galaxies themselves may also contribute to
the sample of stellar clusters. Since their expected, average absolute
magnitude is M$_\mathrm{K} \approx -10$~mag, they will not represent a major
contribution. The mean color indices for 31 Milky Way clusters, including all
the Messier plus Omega Cen and 47 Tuc, are $<$H-K$>$ = 0.24$\pm$0.06 and
$<$J-H$>$ = 0.69$\pm$0.09, which are just above the stellar sequence (see
Fig.~\ref{all-ccd}).

\begin{figure}
  \resizebox{\hsize}{!}{\includegraphics{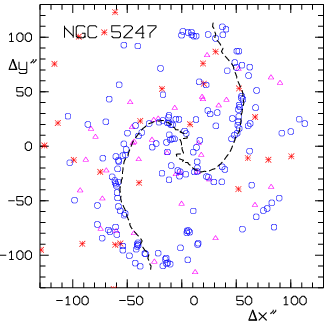}}
  \caption[]{Location of {\sl SExtractor} sources identified in the NGC~5247
    field using the same symbols as in Fig.~\ref{sext}. Coordinates are given
    in arcsec relative to the center of the galaxy in the plane of its
    disk. The dashed lines show the locations of the spiral arms as determined
    from the m=2 Fourier component of the azimuthal K-band intensity
    profiles. }
  \label{n5247xy}
\end{figure}

The spatial distribution of the three source categories, differentiated
according to the {\sl SExtractor} criterion, is shown in a projection
corresponding to that of the disk of NGC~5247 (see Fig.~\ref{n5247xy}) where
the locations of the spiral arms in the K-band are also indicated by dashed
lines as defined by the azimuthal phase of the m=2 Fourier component. Position
angle PA = 110\degr\ and inclination angle IA = 34\degr\ for NGC~5247 were
derived from the K-band image as described by \citet{grosbol04} where the
value of PA differs significantly from the estimate of 20\degr\ given by RC3.
Whereas sources with cs$>$0.2 are uniformly distributed over the field, the
more diffuse sources are strongly concentrated in the arm regions.  This
suggests that most of these sources are complexes of young stellar clusters,
although some of the larger ones could also be background galaxies.  Sources
with an area of more than 100 pixels were excluded to reduce the influence of
background galaxies and complex starforming regions, which may be grouped
together by the detection algorithm. For NGC~5247, this limit corresponds to a
circle with a diameter of 265~pc (i.e. 3.3\arcsec).

\section{Color distribution}
\begin{figure}
  \resizebox{\hsize}{!}{\includegraphics{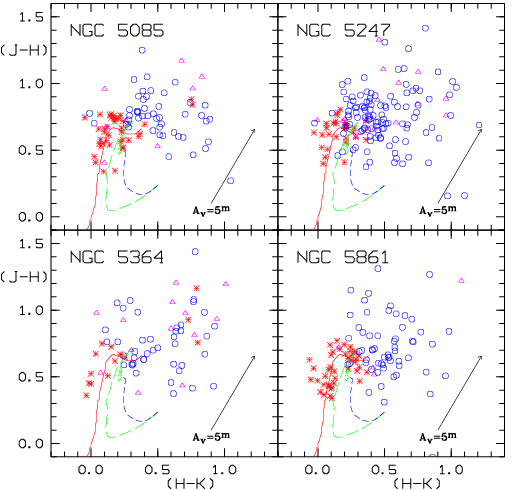}}
  \caption[]{Color-color diagrams of objects in the NGC~5085, NGC~5247,
    NGC~5364, and NGC~5861 fields using the NIR colors (H-K) and (J-H).
    Symbols are the same as in Fig.~\ref{sext}. The fully drawn line indicates
    the location of the local stellar main sequence.  The long and short
    dashed lines show the evolution of a stellar cluster using instantaneous
    and continuous SB99 models, respectively.  The reddening vector at the
    lower right indicates the offset due to 5~mag of visual extinction.
      Typical uncertainties for sources around K=18 mag are 0.2 mag in each
      color index.}
  \label{all-ccd}
\end{figure}

The (H-K)--(J-H) color-color diagrams could be calculated for NGC~5247,
NGC~5086, NGC~5364 and NGC~5861 and are shown in Fig.~\ref{all-ccd}.  To
analyze the diagrams, several SB99 models were computed with upper mass limit
M$_\mathrm{u}$ = 30~M$_{\odot}$, a Salpeter IMF, and solar metallicity.
Models with both continuous or bursty star formation were calculated up to
40~Myr. In Fig.~\ref{all-ccd}, representative evolutionary tracks for one
continuous and one instantaneous model are drawn with the standard
reddening vector using a galactic absorption law.  Both evolutionary tracks
start around (0.5, 0.2) and evolve to bluer (J-H) and (H-K) colors. After the
appearance of the red super-giants (RSG), they redden again in (J-H), while
(H-K) almost remains the same irrespective of the model type.

Extended sources occupying the upper left or lower right corners of the
color-color diagrams cannot be explained by the models, neither by different
star formation histories nor by standard extinction.  Composed colors with
(J-K)$\geq$2 can be explained by the presence of C-stars \citep{dottori05},
which could indicate that such clusters are older than 1~Gyr.  The majority of
the condensations may well be young stellar complexes in different stages of
evolution with different degrees of internal reddening, from zero age HII
regions on, up to 10-20~Myr old complexes. The position of the condensations
with regard to the models indicates that a substantial amount of dust
(i.e. A$_\mathrm{V}$ in the range 2-7~mag) accompanies their evolution during
this phase of their lives.  Because the measure of dust absorption relies on
colors integrated over an entire complex, significant internal variations are
expected where some inner parts may even be obscured in the K-band.  The
outliers amount to a few percent of the samples, with NGC~5247 having slightly
more, and may also represent gross errors due to crowding or matching of
sources from different bands.

\begin{figure}
  \resizebox{\hsize}{!}{\includegraphics{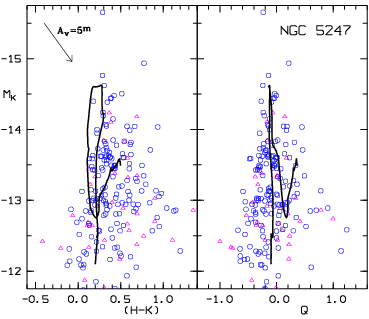}}
  \caption[]{Color-magnitude diagrams of all extended sources in the NGC~5247
    field. The absolute K magnitudes, M$_\mathrm{K}$, were computed from the
    aperture magnitudes assuming a distance of 16.7 Mpc. Diagrams for both
    (H-K) and the `reddening corrected' index Q are shown. An evolution track
    for a bursty SB99 model with a total mass of 10$^4$~M$_{\odot}$ is
    indicated by the line where the tracks start at around (H-K) = 0.5~mag and
    Q = 0.4~mag for a zero age cluster and end at an age of 40~Myr. The
    reddening vector in the upper left corner indicates the shift for a visual
    extinction of 5 mag.  Typical errors for objects with M$_\mathrm{K}$ = -13
    mag are 0.1 mag and 0.2 mag for M$_\mathrm{K}$ and (H-K),
    respectively. Same symbols as in Fig.~\ref{sext} are used. }
  \label{n5247-cmd}
\end{figure}

The color-magnitude diagrams for NGC~5247 are given in Fig.~\ref{n5247-cmd}
where M$_\mathrm{K}$ is shown as a function of (H-K) and a reddening-corrected
index Q = (H-K) - 0.59$\times$(J-H). This reddening correction corresponds to
standard galactic dust extinction for a screen model \citep{rieke85} and will
take into account neither different dust properties nor complex dust-star
geometry in starforming regions \citep{witt92,pierini04}.  The evolution
track of a typical SB99 model with instantaneous star formation and total mass
of 10$^4$~M$_{\odot}$ is overlayed for reference in
Fig.~\ref{n5247-cmd}. Typical errors in (H-K) and Q are in the range of
0.2-0.3~mag for M$_\mathrm{K}$=-13~mag.  The (H-K) colors scatter redwards of
the SB99 model tracks (except for some faint sources with larger errors)
indicating strong attenuation by dust even at NIR wavelengths.  In the
M$_\mathrm{K}$--(H-K) diagram, all the points to the right of the evolutionary
track can be explained with reddening (see reddening arrow in
Fig.~\ref{n5247-cmd}) which is not the case for the `reddening free'
diagram. This most probably indicates that the screening extinction model
is not correct, and dust is mixed with gas and stars in a more complex way in
most of these cases \citep{witt92}. Sources with (H-K)$<$0.0 are too blue
compared to the models and cannot be explained by dust reddening. These
objects can be blue background galaxies, quasars, or sources with
significant non-thermal radiation.  The majority of extended sources
(cs$<$0.8) in NGC~5247, considering problems in reddening corrections, are
compatible with stellar clusters or complexes with total masses up to
5$\times$10$^4$~M$_{\odot}$ using bursty SB99 models (see
Fig.~\ref{n5247-cmd}).  Continuous models give similar results with slightly
higher Q values.  It is not possible to distinguish between the two models
based only on broad band colors but spectroscopic studies suggest a continuous
formation is more likely \citep{grosbol06}.

\begin{figure}
  \resizebox{\hsize}{!}{\includegraphics{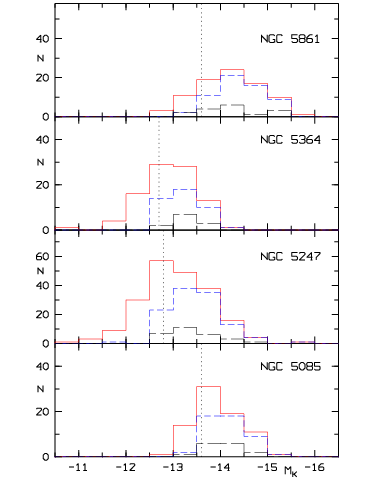}}
  \caption[]{Histograms of the absolute K-band magnitude M$_\mathrm{K}$ of
    objects detected in the images of NGC~5085, NGC~5247, NGC~5364, and
    NGC~5861. The fully drawn line shows the distribution of all extended
    objects.  The instrumental magnitude limit of -4~mag (i.e. with errors
    $<$0.1~mag) is indicated by a vertical dashed line. The histogram with
    short dashed line includes only diffuse objects (cs$<$0.2) brighter than
    this limit, while the long dashed line shows the youngest of these objects
    defined as having 0.1$\leq$Q. }
  \label{mk-hist}
\end{figure}

The distribution of absolute K-band magnitudes, M$_\mathrm{K}$, for extended
sources is shown in Fig.~\ref{mk-hist} for the four galaxies with JHK
photometry.  For each galaxy, the histograms for all extended sources
identified are given with an indication of the limiting magnitude for which
the estimated error is 0.1~mag. Similar diagrams for the full sample of
galaxies are given in the appendix available on-line.  In Fig.~\ref{mk-hist},
the distribution of diffuse sources (cs$<$0.2) and that of the youngest of
these objects (0.0$\leq$Q) are also displayed.  There is a significant spread
in the distribution of the brightest knots, which range up to almost -16~mag
in M$_\mathrm{K}$.  The material is not deep enough to show the full shape or
size of the distributions.  Thus, it is not possible to say whether the
variation of the brightest knots is real or just associated to the population
size in the galaxy.  Although a falling frequency at the faint end of the
distributions is suggested for some galaxies (see e.g. NGC~1365, NGC~2935, and
NGC~5861), the limiting magnitude makes it impossible to make a general
statement on the faint end of the distribution of clusters and complexes.

\section{Spatial distribution}
\begin{figure}
  \resizebox{\hsize}{!}{\includegraphics{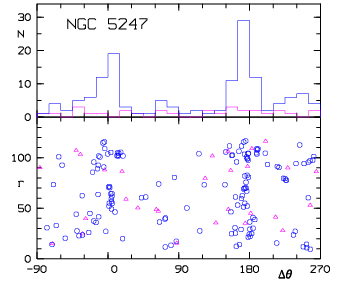}}
  \caption[]{Distribution of extended sources relative to the main
    two-armed spiral pattern in NGC~5247.  The lower panel shows linear radial
    distance, in arcsec, as a function of the azimuthal phase difference
    $\Delta\theta$, in degrees, from the K-band intensity maximum defined by
    the m=2 Fourier component.  The top diagram displays the histogram of the
    distribution. Symbols have the same meaning as in Fig.~\ref{sext}.}
  \label{dth-hist}
\end{figure}

It is of interest to look at the locations of the knots relative to the spiral
pattern in the host galaxy to find possible relations between the formation of
massive stellar clusters and spiral perturbations.  As an example, the spatial
distribution of extended objects relative to the average, two-armed spiral
perturbation in NGC~5247 is shown in Fig.~\ref{dth-hist}.  It gives the linear
radius of the knots as a function of their phase difference $\Delta\theta$
from the m=2 intensity maximum as determined by a Fourier analysis of the
azimuthal K-band intensity variation in the disk.  The histogram of phase
differences is displayed in the top diagram where compact and diffuse sources
are shown separately.  Similar diagrams for all galaxies in the sample are
available in the online material (see Figs.~\ref{fig-a1}-\ref{fig-a9}).

While the compact sources have a uniform distribution, the diffuse objects are
concentrated in the spiral arms (i.e. $\Delta\theta$ = 0\degr and 180\degr)
but still with a significant number of knots in the inter-arm regions.  In
this and several other grand-design galaxies, one can observe average phase
offsets between the knots in the arms and the associated K-band spiral
intensity maximum.  Furthermore, differences in pitch angle between the string
of knots along an arm and that of the K-band spiral may be seen in some
galaxies (e.g. NGC~157, NGC~1566, NGC~2997, NGC~3627, and NGC~5247).  This may
be caused by systematic, radial offsets of large-scale shocks in the gas
relative to the spiral potential minimum \citep{gittins04}, combined with the
drift of material through the arms in a density wave scenario \citep{lin64,
  roberts69a}.  Although these offsets could provide information on the
pattern speed of an underlying density wave, reliable estimates cannot be
obtained without age information on the individual condensations.

When the diffuse sources are concentrated in arms, they seldom have an even
distribution between the two arms.  This suggests that perturbations other
than bisymmetric ones are affecting the formation of the detected
condensations.  A scenario in which the perturbation and the perturbed
material show different temporal phases or trigger an m=1 perturbation, like
in \citet{junqueira96}, may well also be at work.  Another option would be
that the asymmetry is linked to the generation and destruction of molecular
gas clouds in spiral arms if the time scale for these processes \citep{wada99}
is comparable to that with which the gas encounters the arms.

On inspection of both visual images in \citet{SB94} and our K-band maps, we
find that 15 of the 22 grand-design spirals in our sample show a significant
concentration of bright knots in their spiral arms.  The remaining 7
grand-design galaxies show little or no concentration. They have relatively
weak two-armed spiral perturbations, except NGC~895, which also has a
relatively high number of interarm knots and therefore a smaller arm-interarm
difference.  Of the 24 galaxies classified as multiple armed spirals, only
NGC~3389 has a significant increase of knots in one of its arms.  This galaxy
is a member of the Leo group and may have experienced recent interactions.

\begin{figure}
  \resizebox{\hsize}{!}{\includegraphics{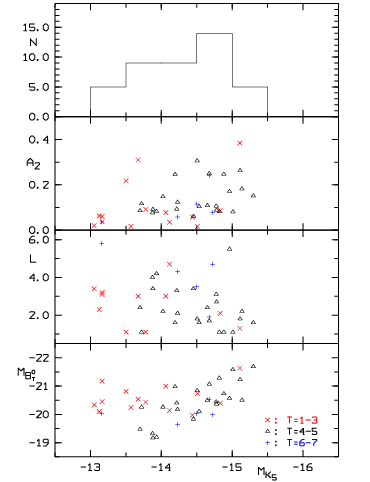}}
  \caption[]{Total absolute B-band magnitude \MBot, luminosity class L, and
    relative amplitude of the two-armed spiral perturbation A$_\mathrm{2}$ as
    functions of the absolute K-band magnitude \MK5\ of the 5$^\mathrm{th}$
    brightest extended sources in the galaxies.  Different symbols indicate
    ranges of morphological type, T, for the galaxies with ($\times$) for
    1$\leq$T$\leq$3, ($\triangle$) for 4$\leq$T$\leq$5, and (+) for
    6$\leq$T$\leq$7.  }
  \label{mk10-dist}
\end{figure}

\section{Statistics of knots}
Characterizing the distributions of young stellar clusters and complexes
in the individual galaxies, only knots within the disks and with cs$<$0.2 were
considered.  The main problems are the contamination by older stellar clusters
and background galaxies and the limiting magnitude of around K = 19~mag, which
does not allow examination of the faint tail of the distributions.  As an
indicator of the most luminous stellar complex in a galaxy, the quantity
\MK5\ was defined as M$_\mathrm{K}$ of the 5$^\mathrm{th}$ brightest
knot.  Especially elliptical galaxies may be difficult to distinguish
morphologically from stellar clusters or complexes.  Assuming that
20-30\% of the background galaxies are elliptical galaxies, counts presented
in Sect.~\ref{sec-sources} suggest less than 6 of them brighter than the
limiting magnitude of this sample (i.e. K~=~19~mag) would be found in an area
of 5 arcmin$^2$ corresponding to the typical disk.  Thus, the choice of the
5$^\mathrm{th}$ brightest knot in the definition of \MK5\ will avoid a strong
bias due to such background galaxies with bright apparent magnitudes.

The distribution of \MK5\ is shown in Fig.\ref{mk10-dist} where correlations
with absolute blue magnitude \MBot, luminosity class L, and amplitude, A$_2$,
of the two-armed spiral pattern for the host galaxy are also given.  The
values of \MK5\ and A$_2$ are listed in Table~\ref{tab2}. While absolute
magnitude and luminosity class were taken from RC3, the amplitudes A$_2$ were
derived from the azimuthal m=2 Fourier component of the K-band maps as
described by \citet{grosbol04}. The overall distribution of \MK5\ ranges from
the detection limit around -13.0~mag (depending on distance modulus) to a
maximum close to -15.5~mag where early-type spirals occupy the fainter range.
For later-type spirals (T$>$3), \MK5\ shows a clear correlation with the
absolute magnitude of the host galaxy.  Generally, brighter values of
\MK5\ are found for galaxies with luminosity class I-II and stronger spiral
perturbations.

\begin{figure}
  \resizebox{\hsize}{!}{\includegraphics{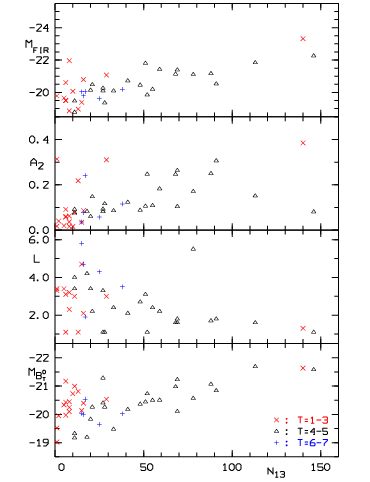}}
  \caption[]{Total absolute B-band magnitude \MBot, luminosity class L,
    relative amplitude of the two-armed spiral perturbation A$_\mathrm{2}$,
    and absolute magnitude of far-infrared flux M$_\mathrm{FIR}$ as functions
    of the number of extended sources \N13\ with the absolute K-band magnitude
    M$_\mathrm{K} <$-13~mag. Symbols are the same as in Fig.~\ref{mk10-dist}.
  }
  \label{n13-dist}
\end{figure}

It is impossible to get a true measure of the population size since the
present sample of images is too shallow to include the faint end of the
stellar cluster/complex population.  The detection limit
M$_\mathrm{K}^\mathrm{l}$, in absolute magnitude, was defined by an
instrumental magnitude of -4~mag corresponding on average to
M$_\mathrm{K}$~=~-13~mag, depending on the distance modulus of the individual
galaxies (see Table~\ref{tab2}). The number of diffuse objects, \N13, with
M$_\mathrm{K}$$<$-13~mag was used to indicate the size of the population of
knots. The values of \N13\ are listed in Table~\ref{tab2}, while
Fig.~\ref{n13-dist} shows the relations to \MBot, L, A$_2$, and absolute
magnitude of far-infrared flux M$_\mathrm{FIR}$ as given by RC3. It is noted
that \MK5\ and \N13\ will correlate if all galaxies have the same distribution
function of knots since a larger population will also have a brighter \MK5
magnitude.

For late-type spirals, both \MK5\ and \N13\ correlate well with the absolute
luminosity of the host galaxy and may be related to the total population size
of young complexes in these galaxies.  The early-type spirals have few
complexes but a wide spread in the luminosity of their brightest member
suggesting that another mechanism beside the population size is important for
the upper brightness limit of the knots.

Galaxies in luminosity class I-II have, on average, more luminous knots, as
measured by \MK5, than those in higher classes as expected from the definition
\citep{bergh60}.  A similar trend is seen for \N13\ where higher luminosity
classes typically have fewer knots than the classes I-II. 

The number of knots correlates with the K-band amplitude, A$_2$, of the
two-armed spiral pattern, whereas galaxies with \MK5$>$-14 have weak spiral
perturbations.  The two exceptions, NGC~3351 and NGC~3627, both have strong
bars that may have biased the estimate of A$_2$.  These relations could
suggest a threshold effect in the sense that only galaxies with strong spiral
perturbations (i.e. 0.1$<$A$_2$) are able to form many and bright stellar
complexes. This may be related to the ability of spiral large-scale shocks
to compress giant molecular clouds \citep{elmegreen81a}.

The magnitude \MK5\ of the brightest knot in a galaxy does not depend on its
far-infrared flux, M$_\mathrm{FIR}$, whereas the population size
\N13\ increases with increasing flux.  Thus, the amount of dust in a galaxy is
likely to influence the number of clusters or complexes formed but not
their luminosity.

\begin{figure}
  \resizebox{\hsize}{!}{\includegraphics{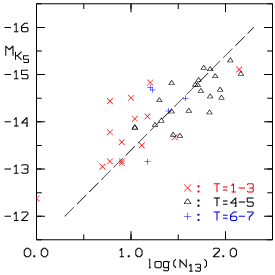}}
  \caption[]{Absolute K-band magnitude \MK5\ of the 5$^\mathrm{th}$
    brightest knot as a function of the logarithm of the number of knots
    \N13\ with M$_\mathrm{K}$$<$-13~mag.  The morphological type of the parent
    galaxies is indicated by different symbols as in Fig.~\ref{mk10-dist}.
    The dashed line indicate a power law relation with a slope of -2.  }
  \label{ln13-mk5}
\end{figure}

Restricting the sample to later type spirals with T$>$3, one may
derive the following two correlations:
$$ \log(\mathrm{N_{13}}) = -0.35(\pm0.06) \times \mathrm{M_{B^0_T}} -
  5.6(\pm0.2) $$ and
$$ \log(\mathrm{N_{13}}) = 1.7(\pm0.6) \times \mathrm{A_2} - 0.17(\pm0.04)
  \times \mathrm{M_{FIR}} + 3.2(\pm0.2) $$
where the dependency on the amplitude of the spiral perturbation is marginal
due to the low values of A$_2$ for the 2 brightest spirals in the sample.

The logarithmic sample size, \N13, and the magnitude of the 5$^\mathrm{th}$
brightest knot given by \MK5\ correlate well as seen in Fig.~\ref{ln13-mk5}.
Although a formal regression gives a shallower slope, the distribution is
compatible with a power law with an index of -2 as found for most luminosity
functions of young clusters \citep[see e.g.][]{whitmore99, whitmore04}.  The
significant spread in the limiting absolute magnitude of the sample will
affect the completeness and bias the distribution.  A general correlation
between sample size and brightest member is expected as a trivial sampling
effect of continuous distribution functions.

\begin{table}
  \caption[]{Statistics for K-band knots in individual galaxies. }
 \label{tab2}
 \begin{tabular}{lrrrrrrrr}
   \hline\hline
   Galaxy  & \multicolumn{1}{c}{A$_\mathrm{2}$} &
   M$_\mathrm{K}^\mathrm{l}$ & \multicolumn{1}{c}{\MK5} &
   \multicolumn{1}{c}{\N13} & \multicolumn{1}{c}{M$_\mathrm{c}$} &
   \multicolumn{1}{c}{$\delta$M} & \multicolumn{1}{c}{SFR} \\ \hline 
   NGC~157  & 0.26 & -13.5 & -15.1 &   69 & -18.9 &  5.4 & 0.73 \\
   NGC~210  & 0.09 & -13.4 & -13.8 &    6 & -16.2 &  7.4 & 0.06 \\
   NGC~289  & 0.15 & -13.2 & -14.0 &   21 & -17.3 &  6.4 & 0.17 \\
   NGC~578  & 0.11 & -13.4 & -14.6 &   55 & -18.4 &  4.6 & 0.47 \\
   NGC~685  & 0.08 & -12.8 & -13.9 &   11 & -16.5 &  5.6 & 0.09 \\
   NGC~895  & 0.24 & -14.1 & -14.7 &   17 & -17.8 &  5.5 & 0.27 \\
   NGC~1068 & 0.06 & -11.3 & -13.1 &    8 & -17.1 &  6.5 & 0.14 \\
   NGC~1084 & 0.11 & -12.9 & -14.8 &   51 & -18.5 &  5.0 & 0.51 \\
   NGC~1087 & 0.17 & -13.2 & -15.0 &   78 & -18.8 &  4.3 & 0.72 \\
   NGC~1187 & 0.12 & -12.8 & -14.2 &   41 & -17.9 &  5.3 & 0.29 \\
   NGC~1255 & 0.09 & -13.3 & -14.2 &   27 & -17.4 &  5.5 & 0.19 \\
   NGC~1300 & 0.25 & -13.1 & -14.9 &   52 & -18.4 &  5.2 & 0.49 \\
   NGC~1350 & 0.08 & -13.4 & -14.1 &   11 & -16.7 &  7.8 & 0.10 \\
   NGC~1357 & 0.06 & -13.7 & -14.4 &    6 & -16.5 &  7.5 & 0.08 \\
   NGC~1365 & 0.38 & -13.1 & -15.1 &  140 & -19.5 &  5.6 & 1.36 \\
   NGC~1371 & 0.04 & -13.0 &    -  &    2 & -14.9 &  8.7 & 0.02 \\
   NGC~1398 & 0.22 & -12.8 & -13.5 &   13 & -16.8 &  7.6 & 0.11 \\
   NGC~1425 & 0.04 & -12.9 & -13.2 &    8 & -16.0 &  7.2 & 0.05 \\
   NGC~1566 & 0.25 & -12.9 & -14.7 &   88 & -18.8 &  5.5 & 0.67 \\
   NGC~2775 & 0.02 & -12.7 & -13.6 &    8 & -16.2 &  7.9 & 0.06 \\
   NGC~2855 & 0.01 & -13.5 &    -  &    2 & -15.0 &  9.0 & 0.02 \\
   NGC~2935 & 0.09 & -13.8 & -14.8 &   16 & -17.7 &  6.4 & 0.25 \\
   NGC~2997 & 0.25 & -12.2 & -14.2 &   68 & -18.7 &  4.8 & 0.64 \\
   NGC~3351 & 0.31 & -11.4 & -12.4 &    1 & -16.0 &  6.9 & 0.05 \\
   NGC~3368 & 0.02 & -11.8 & -13.1 &    5 & -16.4 &  7.3 & 0.08 \\
   NGC~3370 & 0.09 & -12.7 & -13.9 &   11 & -16.5 &  5.3 & 0.08 \\
   NGC~3389 & 0.08 & -12.7 & -13.9 &   18 & -16.9 &  4.7 & 0.12 \\
   NGC~3621 & 0.04 & -10.9 & -13.2 &   15 & -18.2 &  4.2 & 0.42 \\
   NGC~3627 & 0.31 & -11.3 & -13.7 &   29 & -18.5 &  5.2 & 0.54 \\
   NGC~3810 & 0.09 & -12.1 & -13.7 &   33 & -17.5 &  5.1 & 0.20 \\
   NGC~4030 & 0.10 & -13.0 & -14.5 &   69 & -18.5 &  5.4 & 0.54 \\
   NGC~4321 & 0.08 & -13.2 & -15.0 &  146 & -19.5 &  5.4 & 1.31 \\
   NGC~4535 & 0.15 & -13.7 & -15.3 &  113 & -19.7 &  4.8 & 1.55 \\
   NGC~4579 & 0.06 & -13.1 & -13.2 &    6 & -15.3 &  9.5 & 0.03 \\
   NGC~4941 & 0.02 & -12.4 &    -  &    1 & -14.2 &  8.3 & 0.01 \\
   NGC~4965 & 0.08 & -14.0 & -14.7 &   16 & -17.6 &  4.6 & 0.22 \\
   NGC~4981 & 0.06 & -13.5 & -14.5 &   20 & -17.5 &  5.8 & 0.21 \\
   NGC~5085 & 0.09 & -13.6 & -14.8 &   48 & -18.4 &  5.5 & 0.49 \\
   NGC~5134 & 0.03 & -13.3 & -14.1 &   15 & -17.0 &  6.5 & 0.13 \\
   NGC~5247 & 0.31 & -12.8 & -14.5 &   91 & -18.8 &  4.6 & 0.68 \\
   NGC~5364 & 0.12 & -12.7 & -13.7 &   28 & -17.4 &  5.6 & 0.18 \\
   NGC~5668 & 0.06 & -13.3 & -14.2 &   25 & -17.5 &  4.8 & 0.20 \\
   NGC~5861 & 0.18 & -13.6 & -15.1 &   59 & -19.0 &  4.5 & 0.81 \\
   NGC~6384 & 0.08 & -13.5 & -14.8 &   27 & -17.9 &  6.3 & 0.31 \\
   NGC~7213 & 0.02 & -13.9 & -14.5 &   10 & -17.5 &  7.3 & 0.22 \\
   NGC~7418 & 0.12 & -13.1 & -14.5 &   38 & -17.8 &  5.1 & 0.29 \\
  \hline   
\end{tabular}
\end{table}

\section{Discussion}
The total absolute magnitude M$_\mathrm{c}$ of the population of young
  clusters and complexes was estimated by integrating the luminosity of all
diffuse sources (cs$<$0.2) in the individual galaxies. The magnitude
difference $\delta$M between this total luminosity of the complexes and
the parent galaxy is listed in Table~\ref{tab2}.  This suggests that at most a
few percent of the K-band flux of a normal spiral galaxy originates from very
young, starforming complexes.

Without detailed color information on the individual sources, it is difficult
to estimate the SFR associated with the bright K-band knots.  A crude
calculation can be done by assuming that the young complexes have an age
around 10~Myr.  Using SB99 models with a standard Salpeter IMF and solar
abundances, such complexes with an SFR of 1~M$_{\odot}$ yr$^{-1}$ would
have an absolute magnitude M$_\mathrm{K} \approx -19.2$~mag. The SFR can then
be estimated and is listed in Table~\ref{tab2}.  These values may have errors
of almost an order of magnitude due to the strong luminosity evolution around
6~Myr.

The possible correlation of the knots presented in this paper with H$\alpha$
emitting complexes, such as those used by SINGG \citep{meurer06, hanish06},
remains an open issue. The NIR emitting knots and the H$_\alpha$ emitting
complexes could present two different moments in the process of star
formation, or they could be different parts of the same complexes.  The
detection of NIR-emitting knots in our sample represents an intermediate
wavelength between SINGG and the SONGG survey proposed for the Spitzer
telescope to be observed in 8~$\mu$m.  While SINGG and SONGG look at global
fluxes, we instead try to observe with higher spatial resolution in order to
study the connection to the perturbations that trigger the process of star
formation.  Thus, an identification of NIR-emitting knots and
H$_\alpha$-emitting ones necessitates imagery with the same resolution.

Our sample has only one galaxy, NGC~895, in common with the SINGG survey. The
estimated SFR is compatible within the errors being 40\% higher than that
given by \citet{meurer06}.  There are 5 galaxies (i.e. NGC~3351, NGC~3621,
NGC~3627, NGC~4321, and NGC~4579) with H$_\alpha$-based SFR in
\citet{kennicutt03}.  These estimates are more than an order of magnitude
higher than our values.  It could suggest that we only measure the most
massive complexes representing the very upper range of the IMF, while SFR
derived from H$_\alpha$ are sensitive to the formation of lower mass stars.

Another important issue to be tackled from our results is the behavior of dust
during the NIR knots' evolution. In fact, comparison with SB99 models for the
four galaxies observed in JHK colors shows that very young NIR knots
($<$2~Myr), as well as the oldest ones (50-70~Myr), approach the models more
than do intermediate age knots. Only at the very end of the model's life are
there some knots that might be considered absorption free
(Fig.~\ref{all-ccd}). This may indicate that processes occurring during the
life of the knots are renewing and/or remodeling the dust content so as to
increase the optical depth.  Contaminating candidates are the beginning of the
RSG phase, as well as other poorly collimated outflows, such as those observed
in Orion IRc2 \citep{Churchwell02}.

\section{Conclusion}
The K-band images of spiral galaxies provide an almost clear view of their
young stellar clusters and starforming complexes that in visual bands are
partly obscured by dust absorption (e.g. by dust lanes in spiral arms).  This
allows an unbiased study, within the resolution and magnitude limits imposed
by the present sample, of the distribution of such complexes in 46 nearby
spiral galaxies.

The (H-K)--(J-H) diagrams for the 4 galaxies, for which JHK photometry is
available, suggest that the majority of diffuse sources (cs$<$0.2) are
starforming complexes that have been reddened with more than 4~mag of visual
extinction.  This is supported by their concentration in arm regions in many
grand-design galaxies.  Since the clusters complexes are unresolved, the
internal attenuation by dust could vary very significantly and may obscure
part of it even in the K-band.  Some background galaxies may be included,
whereas globular clusters are expected to be significantly fainter. Due to the
high extinction and lack of color information for the majority of the sample,
detailed conclusions cannot be made on the star formation process and its
parameters.

Around 70\% (15/22) of grand-design spirals in the sample show significant
concentration of bright, diffuse knots in their arms corresponding to 30\%
(15/46) of the complete sample.  The 7 grand-design systems that do not
display a significant concentration of knots all have weak two-armed spiral
perturbations.  This suggests that only disk galaxies with strong spiral
perturbation (i.e. 0.1$<$A$_2$) will be able to develop a starforming front
associated to its spiral pattern, while galaxies with weaker spirals will have
a more random star formation in their disks.

The brightest clusters/complexes as measured by \MK5, the
5$^\mathrm{th}$ brightest diffuse knot, reach absolute magnitudes of
M$_\mathrm{K} \approx -15.5$~mag but depend on absolute magnitude, type,
luminosity type, and strength of the spiral perturbation of the host
galaxy. Early type spirals have, on average, fainter complexes than
later types.  The brightest complexes are found in more luminous
late-type spirals in luminosity class I-II with strong two-armed spiral
patterns.

The number of clusters and complexes in a galaxy indicated by \N13,
diffuse knots with M$_\mathrm{K} < -13$~mag, shows the same trends as \MK5. In
addition, the number correlates well with the far-infrared flux, suggesting a
dependency on the amount of dust in the galaxy.

Star formation rates were estimated from the total integrated K-band
luminosity of the diffuse knots in each galaxy.  The values range up to
1~M$_{\odot}$ yr$^{-1}$  for grand-design spirals but may only indicate the
formation of the most massive complexes.

\begin{acknowledgements}
  The ESO-MIDAS system and SExtractor were used for the reduction and analysis
  of the data. HD thanks ESO and the Brazilian Council of Research CNPq,
  Brazil, for support.  We would also like to thank the referee,
    Dr. P.~Anders, for his comments and suggestions, which significantly
    improved the paper.
\end{acknowledgements}

\bibliographystyle{aa}
\bibliography{AstronRef}

\Online
\begin{appendix}
\section{Individual galaxies}
\begin{figure}
  \resizebox{\hsize}{!}{\includegraphics{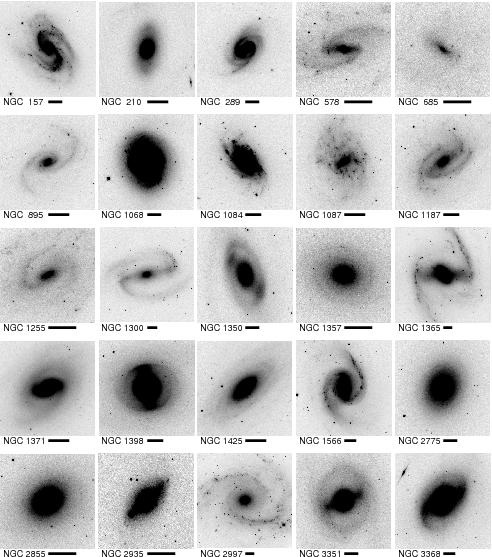}}
  \caption[]{Direct K-band image of the galaxies in the sample.  The scale of
    the image is indicated by the bar which corresponds to 30\arcsec.}
  \label{fig-b1}
\end{figure}

\begin{figure}
  \resizebox{\hsize}{!}{\includegraphics{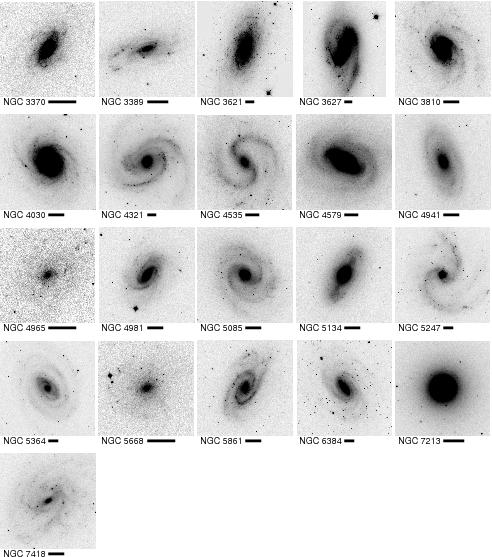}}
  \caption[]{Direct K-band image of the galaxies in the sample.  The scale of
    the image is indicated by the bar which corresponds to 30\arcsec.}
  \label{fig-b2}
\end{figure}

\begin{figure}
  \resizebox{\hsize}{!}{\includegraphics{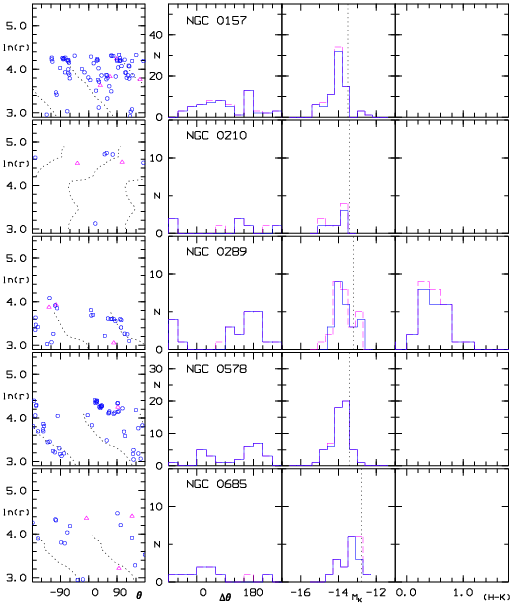}}
  \caption[]{Distribution of extended objects in the fields of NGC~157,
    NGC~210, NGC~289, NGC~578, and NGC~685. For each galaxy, 4 diagrams are
    provided. From left to right: 1) location of extended objects in a
    $\theta-\ln(r)$ diagram, 2) histogram of the azimuthal distances from the
    m=2 spiral component as in Fig.~\ref{dth-hist}, 3) histogram of absolute
    K-band magnitudes, M$_\mathrm{K}$, and 4) histogram of colors (H-K) for
    the galaxies for which both colors are available.  }
  \label{fig-a1}
\end{figure}

\begin{figure}
  \resizebox{\hsize}{!}{\includegraphics{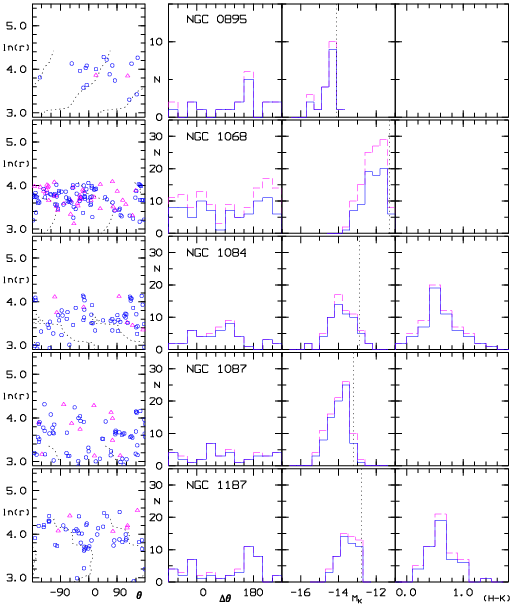}}
  \caption[]{Distribution of extended objects in the fields of NGC~895,
    NGC~1068, NGC~1084, NGC~1087, and NGC~1187. Diagrams are as described in
    Fig.~\ref{fig-a1}}
  \label{fig-a2}
\end{figure}

\begin{figure}
  \resizebox{\hsize}{!}{\includegraphics{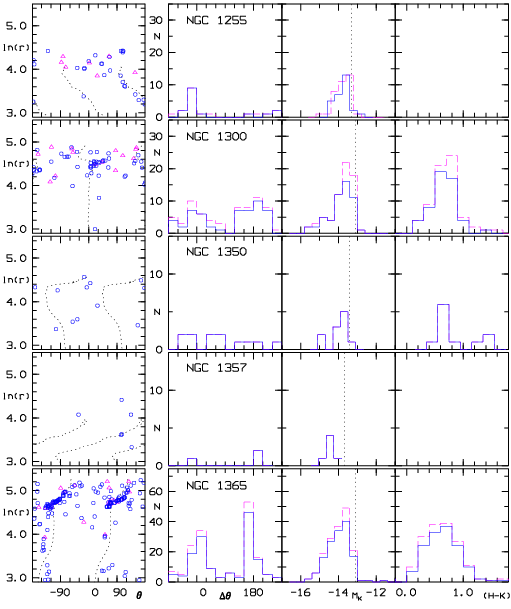}}
  \caption[]{Distribution of extended objects in the fields of NGC~1255,
    NGC~1300, NGC~1350, NGC~1357, and NGC~1365. Diagrams are as described in
    Fig.~\ref{fig-a1}}
  \label{fig-a3}
\end{figure}

\begin{figure}
  \resizebox{\hsize}{!}{\includegraphics{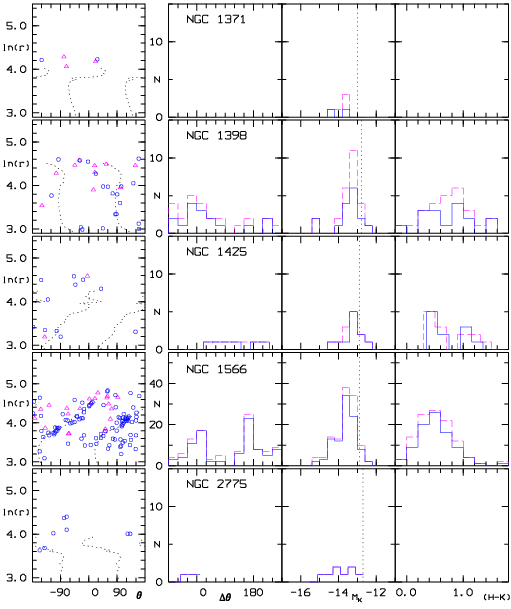}}
  \caption[]{Distribution of extended objects in the fields of NGC~1371,
    NGC~1398, NGC~1425, NGC~1566, and NGC~2775. Diagrams are as described in
    Fig.~\ref{fig-a1}}
  \label{fig-a4}
\end{figure}

\begin{figure}
  \resizebox{\hsize}{!}{\includegraphics{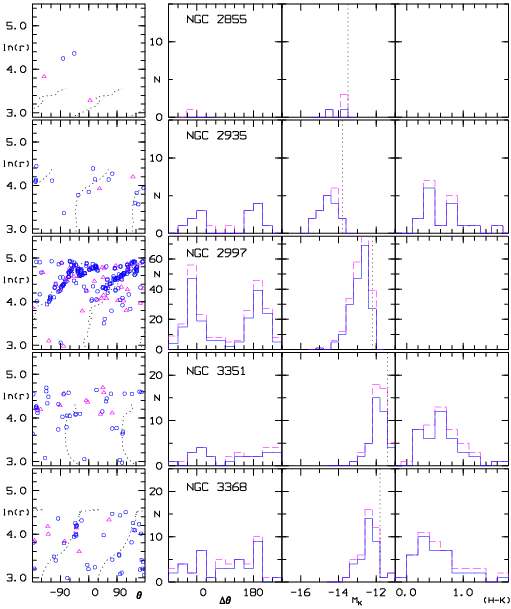}}
  \caption[]{Distribution of extended objects in the fields of NGC~2855,
    NGC~2935, NGC~2997, NGC~3351, and NGC~3368. Diagrams are as described in
    Fig.~\ref{fig-a1}}
  \label{fig-a5}
\end{figure}

\begin{figure}
  \resizebox{\hsize}{!}{\includegraphics{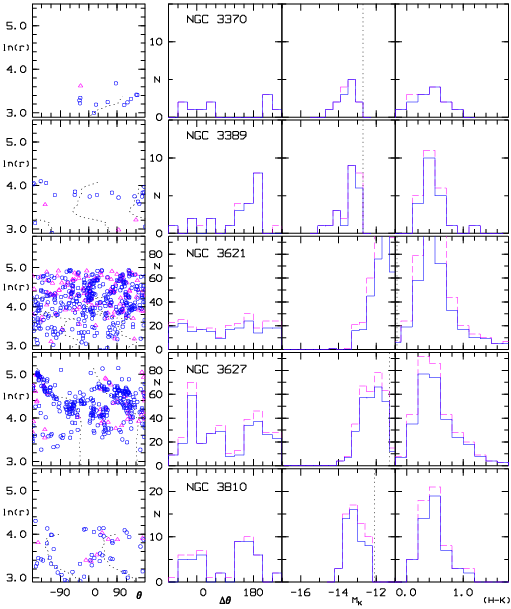}}
  \caption[]{Distribution of extended objects in the fields of NGC~3370,
    NGC~3389, NGC~3621, NGC~3627, and NGC~3810. Diagrams are as described in
    Fig.~\ref{fig-a1}}
  \label{fig-a6}
\end{figure}

\begin{figure}
  \resizebox{\hsize}{!}{\includegraphics{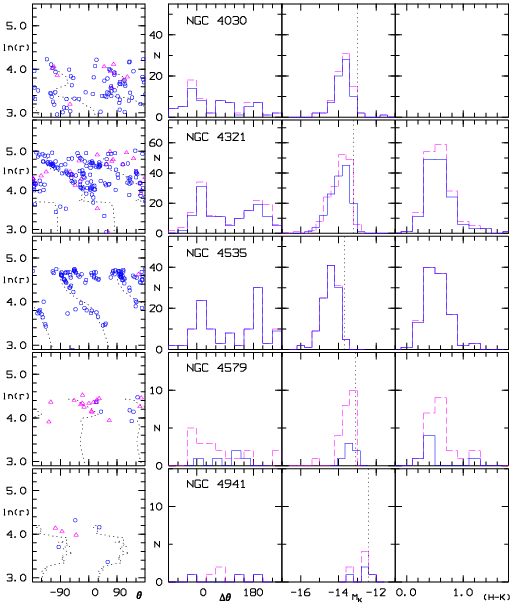}}
  \caption[]{Distribution of extended objects in the fields of NGC~4030,
    NGC~4321, NGC~4535, NGC~4579, and NGC~4941. Diagrams are as described in
    Fig.~\ref{fig-a1}}
  \label{fig-a7}
\end{figure}

\begin{figure}
  \resizebox{\hsize}{!}{\includegraphics{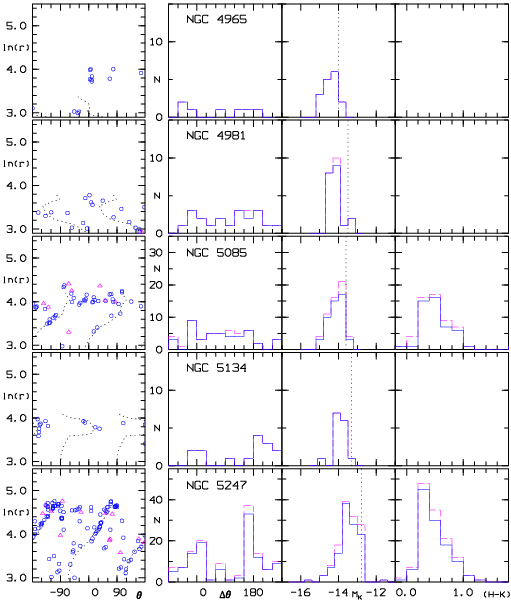}}
  \caption[]{Distribution of extended objects in the fields of NGC~4965,
    NGC~4981, NGC~5085, NGC~5134, and NGC~5247. Diagrams are as described in
    Fig.~\ref{fig-a1}}
  \label{fig-a8}
\end{figure}

\begin{figure}
  \resizebox{\hsize}{!}{\includegraphics{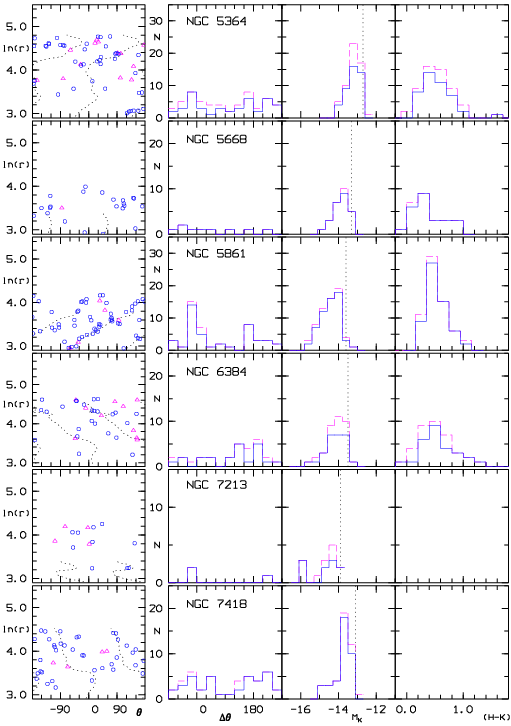}}
  \caption[]{Distribution of extended objects in the fields of NGC~5364,
    NGC~5668, NGC~5861, NGC~6382, NGC~7213, and NGC~7418. Diagrams are as
    described in Fig.~\ref{fig-a1}}
  \label{fig-a9}
\end{figure}

This section provides direct K-band images for the entire sample of galaxies
used to study the statistics of bright knots in spiral galaxies (see
Fig.~\ref{fig-b1}--\ref{fig-b2}). Diagrams of the distribution of extended
sources in the individual galaxies are shown in
Fig.~\ref{fig-a1}--\ref{fig-a9}. For each galaxy, the leftmost plot gives the
locations of knots in a polar $\theta-\ln(r)$ map where the position of the
two-armed spiral pattern, as derived from the phase of the azimuthal m=2
Fourier component, is indicated by dashed lines.  Sources with {\sl
  star\_class} in the range 0.2$<$cs$\leq$0.8 are plotted as ($\triangle$)
while the most diffuse ones with cs$\leq$0.2 are shown as (+).  The next
diagram shows the azimuthal distribution of the knots relative to the spiral
pattern where 0\degr\ and 180\degr\ indicates the maxima of K-band intensity
of the two-armed spiral.  It is noted that only 15 of the galaxies (mainly
grand-design spiral) show a concentration of knots in their arm regions. The
third plot gives the histogram of absolute magnitudes M$_\mathrm{K}$, where the
vertical dotted line indicates the limiting magnitude
M$_\mathrm{K}^\mathrm{l}$, The rightmost shows the distribution of the (H-K)
color index for galaxies when available.  For the histograms, the top, dashed
line indicates the distribution for all knots with cs$\leq$0.8 whereas the
lower full drawn line only includes knots with cs$\leq$0.2.

\end{appendix}

\end{document}